\documentclass[a4paper,twoside]{article}
\usepackage{listings}
\usepackage{courier}

\usepackage{epsfig}
\usepackage{subcaption}
\usepackage{calc}
\usepackage{amssymb}
\usepackage{amstext}
\usepackage{amsmath}
\usepackage{amsthm}
\usepackage{multicol}
\usepackage{pslatex}
\usepackage{apalike}
\usepackage[bottom]{footmisc}
\usepackage{SCITEPRESS}     

\usepackage{comment}
\usepackage{multirow}
\usepackage{tabularx,booktabs}

\begin{document}

\title{Software Vulnerability Prediction Knowledge Transferring Between Programming Languages}

\author{\authorname{Khadija Hanifi\sup{1}\orcidAuthor{0000-0001-7044-3315}, Ramin F Fouladi\sup{1}\orcidAuthor{0000-0003-4142-1293}, Basak Gencer Unsalver\sup{2}\orcidAuthor{0000-0001-9426-8400} and Goksu Karadag\sup{2}\orcidAuthor{0000-0002-4596-0983}}
\affiliation{\sup{1}Ericsson Security Research, Istanbul, Turkey}
\affiliation{\sup{2}Vodafone, Istanbul, Turkey}
\email{{khadija.hanifi, ramin.fuladi}@ericsson.com, {goksu.karadag, basak.gencer}@vodafone.com}}

\abstract{Developing automated and smart software vulnerability detection models has been receiving great attention from both research and development communities. One of the biggest challenges in this area is the lack of code samples for all different programming languages. In this study, we address this issue by proposing a transfer learning technique to leverage available datasets and generate a model to detect common vulnerabilities in different programming languages. We use C source code samples to train a Convolutional Neural Network (CNN) model, then, we use Java source code samples to adopt and evaluate the learned model. We use code samples from two benchmark datasets: NIST Software Assurance Reference Dataset (SARD) and Draper VDISC dataset. The results show that proposed model detects vulnerabilities in both C and Java codes with average recall of 72\%. Additionally, we employ explainable AI to investigate how much each feature contributes to the knowledge transfer mechanisms between C and Java in the proposed model. }

\onecolumn \maketitle \normalsize \setcounter{footnote}{0} \vfill

\section{\uppercase{Introduction}}
\label{sec:introduction}
Developers are concerned with the correctness of the written codes to run in a desired way and meet predefined design specifications. Along with that, they utilize different code analyzing techniques to ensure that the written code is robust enough and free of any weaknesses (called vulnerabilities) that could be exploited by attackers to carry out their malicious activities. Managing software vulnerabilities involves a wide range of code analyzing techniques to enhance the security and to ensure the confidentiality, integrity, and availability of the system. Source code analyzing techniques are usually classified into two main groups; static and dynamic code analysis \cite{palit2021dynpta}. Static code analysis examines the source code without executing and running the code. 
It utilizes pre-defined set of rules and codding standards to analyze how much the code meets these rules and standards. On the other hand, dynamic code analysis entails running code and examining the outcome. 
This involves testing different possible execution paths of the code and examining their outputs. Dynamic code analysis is able to find security issues caused by the program interaction with other system components such as SQL databases or Web services. However, in dynamic code analysis, to build a series of correct inputs for test coverage, a pre-knowledge of the program steps is needed. Although static and dynamic analysis have different approach to find the vulnerabilities, they both suffer from high false positive rate, 
necessitating human expertise to review the results which yields extra time, effort, and cost. 

Many studies in literature have explored methods to decrease the false positive rate of both static and dynamic analysis. A common approach utilized in these studies is the use of machine learning (ML) techniques to train vulnerability detection models \cite{lin2020software}. While ML-based models show potential for improving vulnerability prediction accuracy, they also face two significant challenges. Firstly, the process of addressing various types of vulnerabilities is resource-intensive and requires both software and human expert analysis. This is further complicated by the fact that commercially used source code is subject to intellectual property rights and considered as confidential information by enterprises, making it difficult to obtain access to real labeled vulnerability data. To mitigate this issue, synthetic datasets, such as the NIST Software Assurance Reference Dataset \footnote{https://samate.nist.gov/SARD} and Draper VDISC dataset, are generated to simulate different vulnerability types for specific programming languages. However, these datasets are limited in scope, covering only published vulnerability types for a limited range of programming languages. Secondly, the structure and logic of each programming language are distinct, which results in language-specific feature extraction and varying numbers of features. This makes it challenging to develop a single model that can be used for all programming languages.

In this study, we propose a novel technique to address the challenge of limited vulnerability sample availability in some programming languages. Our approach utilizes available datasets to train a machine learning model for vulnerability detection in a related programming language and applies this knowledge to predict vulnerabilities in other programming languages with limited training samples. The main contributions of our work are:
\begin{itemize}
    \item An automated and interpretable software vulnerability prediction model using machine learning techniques, which can learn from datasets in one programming language and apply its knowledge to predict vulnerabilities in other programming languages.
    \item A code representation method that converts source code into numerical vectors for machine learning analysis, combined with a syntax matching technique for matching related components between Abstract Syntax Trees (ASTs) from different programming languages.
    \item A new method for vulnerability prediction knowledge transfer, which involves customized processing steps for different programming languages and common final steps for vulnerability prediction.
\end{itemize}
We carried out a preliminary experiment to demonstrate our approach using the common vulnerabilities between C and Java programming languages. Our results show that we were able to transfer vulnerability prediction knowledge from C programming language to Java and that our machine learning model was able to make accurate predictions. Additionally, we leveraged explainable AI methods to better understand and verify the correctness of our model.
\vspace{-1.05em}
\section{\uppercase{Related Work}}\label{sec:literature}
Software vulnerability prediction is dramatically rising in popularity in research community and industry, especially among cyber security professionals dealing with vulnerability management and secure software development life cycle (SSDLC). This spike of popularity is due to the rise in cyber security threats which exploit vulnerabilities for their malicious intent. Although several studies exist in the literature for software vulnerability prediction, the use of ML-based approaches are ubiquitous in the literature \cite{hanif2021rise}. In general, two main approaches of metric-based and pattern-based are utilized with ML algorithms for software vulnerability prediction \cite{li2021automated}. 
Metric-based approaches link the software engineering metrics, such as code complexity metrics \cite{kalouptsoglou2022examining,rusen} and developer activity metrics \cite{coskun2022profiling} to software vulnerabilities and use those metrics to train ML models for vulnerabilities prediction. Although metric-based approaches are lightweight to analyze a large-scale program, they suffer from high positive rates. On the other hand, pattern-based approaches improve the efficiency and the automation of software vulnerability prediction. 
Authors in \cite{bilgin2020vulnerability} proposed a method for software vulnerability prediction on the function level for C code. While preserving the structural and semantic information in the source code, the method transforms the AST of the source code into a numerical vector and then utilizes 1D CNN for software vulnerability prediction. 
Authors in \cite{duan2019vulsniper}, used the Control Flow Graph (CFG) and AST as the graph representation and utilized soft attention to extract high-level features for vulnerability prediction. Authors in \cite{zhou2019devign}, proposed a function-level software vulnerability prediction method based on a graph representation that utilize AST, dependency, and natural code sequence information.  

With respect to applying transfer learning for software vulnerability prediction, authors in \cite{ziems2021security} developed various deep learning Natural Language Processing (NLP) models to predict vulnerabilities in the C/C++ source code. They used transfer learning to adapt some pre-trained models for English language such as Bidirectional Encoder Representations from
Transformers (BERT) \cite{devlin2018bert} to be reused for software vulnerability prediction. The performance is promising despite the different structure between English language and the C/C++ programming language. Authors in \cite{lin2019software}, proposed a software vulnerability prediction method based on transfer learning and Long-short Term Memory (LSTM). They used several heterogeneous and cross-domain data sources combined to obtain a general representation of patterns for vulnerability prediction.  

Although pattern-based approaches improve the prediction performance with respect to false positive rate, they are not adaptable and flexible enough to be re-used in other domains rather than the domain they are trained for. TL-based methods address this issue to some extent; however, they are mainly limited to very close domains such as the same or similar programming languages. In this study, we address the aforementioned issues by combing a pattern-based approach with transfer learning to transfer the knowledge acquired for software vulnerability prediction in C source code being used for software vulnerability prediction in Java source code.  
\section{\uppercase{Proposed Method}}
\label{sec:method}  
In this section, we discuss our proposed method to develop an ML-based programming language agnostic software vulnerability prediction model. To this end, we trained a CNN model with software vulnerability samples of C source code, and then, we updated trained model to predict vulnerabilities in Java source code. The main phases of the proposed method are discussed in the following sub-sections.

\subsection{Preprocessing source code}
Extracting discriminating features from source code and representing them as numerical vectors is considered the most critical step in training an ML-based software vulnerability prediction model. Converted numerical representation not only needs to preserve the essential information from the source code, but also needs to preserve the semantic relations between the key elements of the source code. Moreover, extracting the optimal feature set improves the vulnerability prediction performance and facilitates training a generic model. 

In this phase, we explain the steps we follow to convert an input source code into an ML suitable numerical vector that contains the most important features for vulnerability prediction. 

We demonstrate the steps of converting a source code into a numerical vector on the code sample provided below, as it is simple and runnable with both C and Java compilers: 

\begin{small}
\begin{verbatim}
int sum(int a,int b){
    return a+b;
\end{verbatim}
\end{small}

\begin{figure}[htbp]
    \centering
    \includegraphics[width=.85\linewidth]{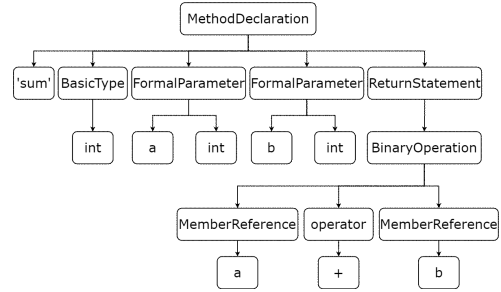}
    \caption{Java AST of "sum" function.}
    \label{fig:ast}
\end{figure}

We convert the input source code into a numerical array through the following steps:

\textbf{Step-1 (Tokenization):} 
In this step, we first make sure that all non-code related additional parts like comments, tabs, and newlines are removed. Then, we convert the main code into a stream of tokens. Each token is represented as a sequence of characters that can be treated as a unit in the grammar of the corresponding programming language. For this, we use two lexers developed explicitly for Java and C source code. The extracted tokens from $sum$ function are as follow:

\begin{lstlisting}
'int': Keyword, 'sum': Identifier, '(': Separator, 'int': Keyword, 'a': Identifier, 'int': Keyword,  'b': Identifier, ')': Separator, '{': Separator,
'return': Keyword, 'a': Identifier, '+': Operator, 'b': Identifier,
';': Separator, 
'}': Separator.
\end{lstlisting}

\textbf{Step-2 (AST Generation, Normalization, and Simplification):}
AST contains syntax and semantic information about the source code, and therefore, it is highly useful for code analysis. In this step, we extract the AST of the tokenized code using two parsers developed explicitly for both Java and C source code. Figure ~\ref{fig:ast} presents the Java AST extracted from $sum$ function.

\begin{figure}[htbp]
    \centering
    \includegraphics[width=.85\linewidth, height=3.5cm]{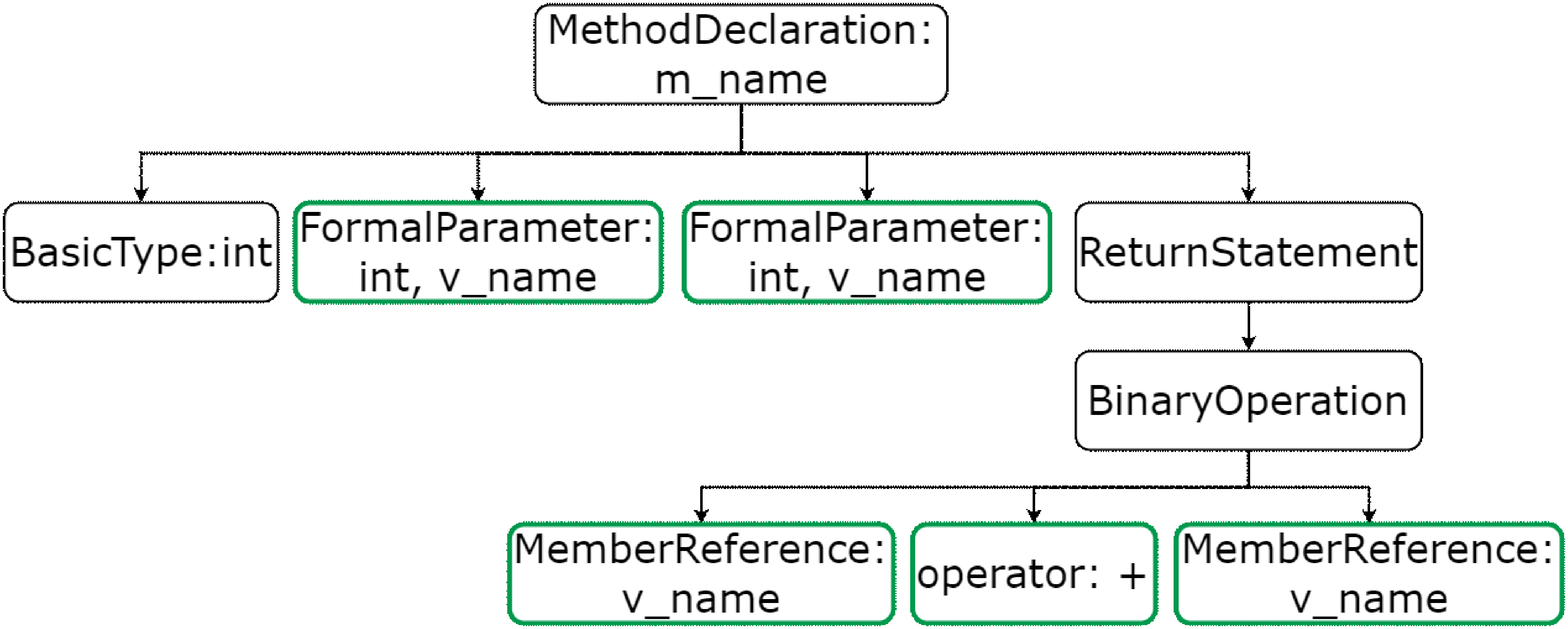}
    \caption{Normalized and simplified Java N-array-AST of "sum" function.}\centering
    \label{fig:normalized_ast}
\end{figure}

Because Java and C languages have distinct structures, there is no common parser for both languages, and thus generated ASTs are different for each language. Also, the AST structure is flexible and each node in the AST can have multiple attributes/children, however, with this structure, it is hard to follow the relationships between AST nodes after converting the AST to an array. Thus, we normalize and simplify C and Java ASTs to ensure that equivalent ASTs will be converted into the same numerical array by the end of the preprocessing phase. We do this by: first, normalizing the AST tokens and replacing all identifier tokens with assigned abstract representation forms. For example, all method names are replaced with "m\_name", and all variable names are replaced with “v\_name”, etc. Then, considering the importance of structural relations between the AST nodes for vulnerability prediction, we simplify the tokens that represent features (leaves in the AST) rather than independent tokens by combining them with their parent token. In Figure ~\ref{fig:normalized_ast} we represent normalized and simplified AST of the function $sum$. 

\begin{figure}[htbp]
    \centering
    \includegraphics[width=.85\linewidth]{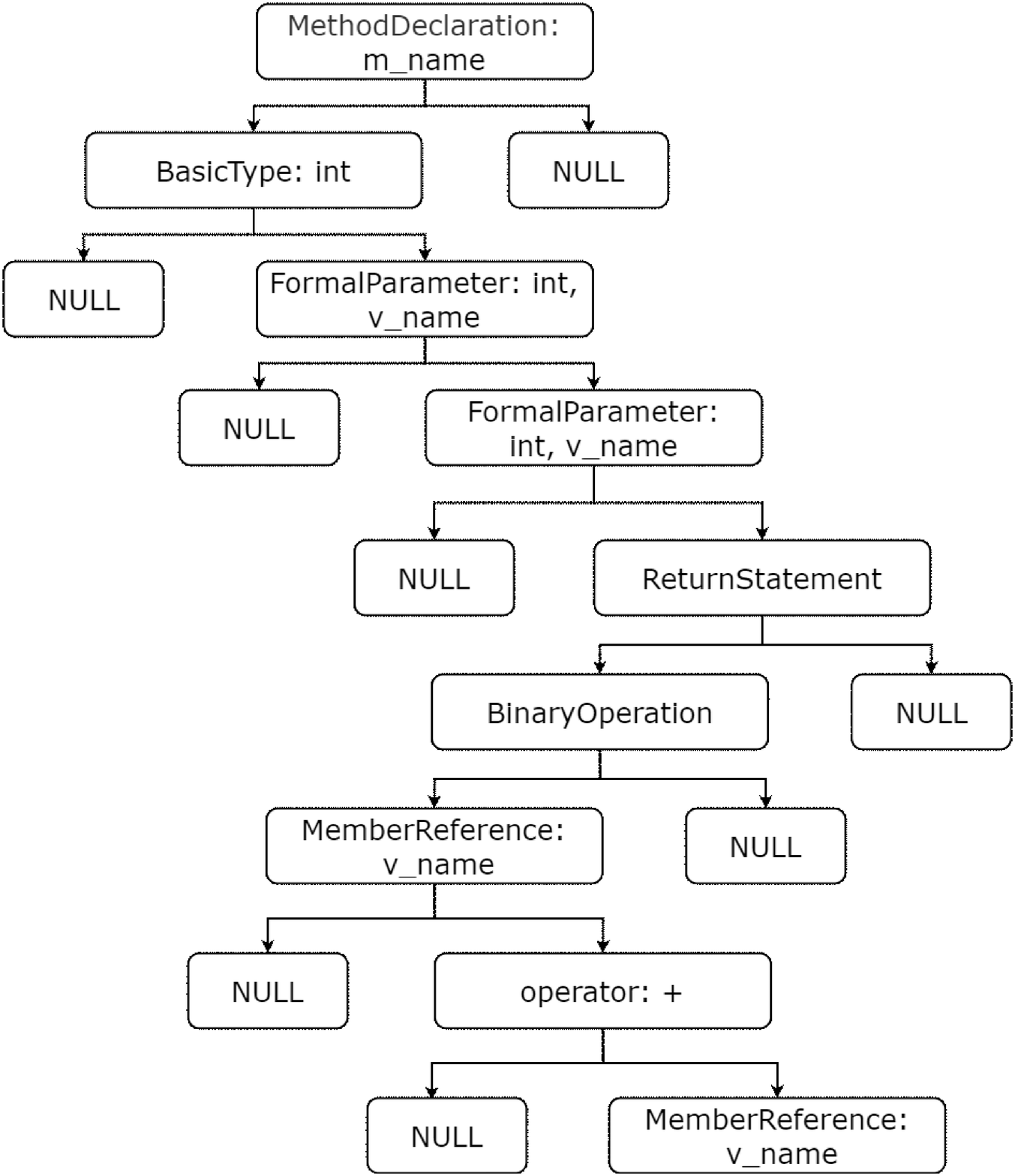}
    \caption{Java binary AST of "sum" function.}
    \label{fig:Java_CB-AST}
\end{figure}

\textbf{Step-3 (Conversion to a Complete Binary AST):}
In this step, we convert the AST from normal tree with unbounded number of children per node into a complete binary tree. By this conversion we not only preserve the relations between nodes but we also unify the AST structure for both C and Java. 
We convert the regular AST (referred as N-array-AST) into a complete binary AST (referred as CB-AST), where all leaves have the same depth, and all internal nodes have exactly two children. The conversion from the N-array-AST to the CB-AST is achieved with respect to the following rules ~\cite{bilgin2020vulnerability}:
\begin{itemize}
    \item The root of the N-array-AST is the root of the CB-AST.
    \item The left-child of a node in the CB-AST is the leftmost child of the node in the N-array-AST.
    \item The right-child of a node in the CB-AST is the right sibling of the node in the N-array-AST.
    \item When the node does not have children, then its left-child is set as NULL.
    \item When the node is the rightmost child of its parent, its right-child is set as NULL
\end{itemize}
The Java CB-AST for $sum$ function is represented in Figure ~\ref{fig:Java_CB-AST}. Additionally, to have a complete tree, NULL children nodes are added until we reach a level when all nodes are NULL.

\textbf{Step-4 (Encoding to Numerical 3-Tuples):}
Since majority of ML algorithms expect numerical inputs, we convert the CB-AST into a numerical vector. To this end, we leverage the encoding method presented in \cite{bilgin2020vulnerability} to convert CB-AST nodes into n-tuples which consists of numerical values. In this study, we used 3-tuple of numbers; the first number represents the type of the token, while second and third numbers store additional information about the nodes.
For example, the token “BasicType: int, 50” is encoded to (8.0, 103.0, 50.0). This representation is considered as a three-dimensional data structure where each dimension holds a value related to an associated token. Notice that the numeric values used for the encoding mechanism are chosen arbitrarily and could be changed as long as different categories take different values. 

\begin{figure}[htbp]
    \centering
    \includegraphics[width=\linewidth]{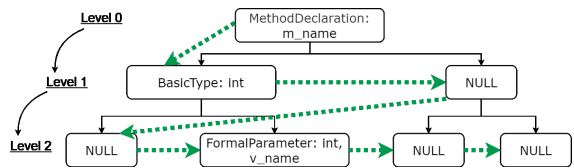}
    \caption{Reading order of the CB-AST nodes.}
    \label{fig:BFS_Java_CB-AST}
\end{figure}

\textbf{Step-5 (Vector Representation):}
After replacing each node in CB-AST with its 3-tuple numeric value, we transfer the CB-AST into a 1D array. To this end, we used Breadth-first search (BFS) algorithm to position the numeric tuples in their corresponding locations in a 1D array. The 1D array is filled by the CB-AST nodes starting from the root node traversing all the neighbor nodes at the present level then moving on to the nodes at the next depth level and so on. In Figure ~\ref{fig:BFS_Java_CB-AST}, we present the BFS algorithm passing through the nodes of the first three levels of the Java CB-AST.

Proposed approach preserves the structural relations in the source code via respective positional relations between elements in the 1D array. Also, with the rich numerical encoding technique (explained in step-4), the semantic information of the AST are preserved and embedded in the 1D array. Another significant advantage of the proposed numerical array representation is that each element in the 1D array acts as a feature, and hence, it can be directly used as an input to an ML model, which facilitates conducting various automated and intelligent code analysis.

\subsection{Syntax matching and knowledge transferring}
\begin{figure}[htbp]
 	\centering
 	\includegraphics[width=0.75\linewidth,]{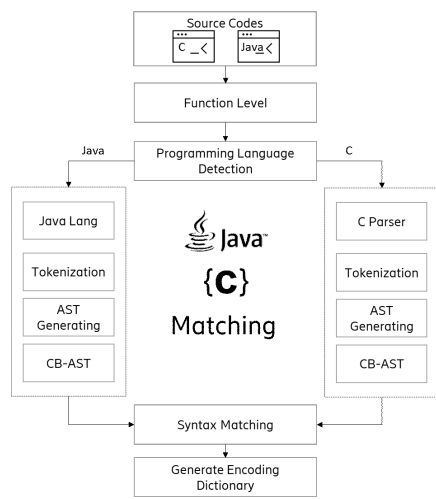}
 	\caption{Steps for matching c and Java tokens.}
	\label{c_Java_matching_steps}
\end{figure}

Analysed tokens are clustered under three main groups: tokens that exist only in C language like CompoundLiteral, tokens that exist only in Java language like ClassDeclaratio, and tokens that exist in both C and Java. Thus, we prepared a dictionary that involves these three categories. However, the tokens that exist in both C and Java are equivalent but different names are assigned due to using different lexers and parser. For example: "FuncDef", "Constant", and "IF" are used by C lexer, whereas, "MethodDeclaration", "BasicType", and "IfStatement" are used by Java lexer. In order to use one model for both Java and C codes, we need to generate the same numerical array for equivalent C and Java codes. Thus, we created a table to match the tokens used by the compiler to construct different statements and expressions in each programming language. We prepared a mapping dictionary following the steps provided in Figure ~\ref{c_Java_matching_steps} to encode equivalent tokens similarly. After generating the CB-AST, tokens are encoded using the prepared dictionary that contains the equivalent tokens of C and Java with their 3-tuples. In Table \ref{tab:CandJavaTokens}, examples of common tokens between C and Java are provided along with their numerical 3-tuples. 

\begin{table}[b]
\fontsize{8}{9}\selectfont
\caption{Examples of common C and Java tokens.}\centering
\begin{center}
\begin{tabular}{|l|l|c|}
\hline
\textbf{C Token} & \textbf{Java Token} & \textbf{Numerical tuples} \\
\hline
  DoWhile   &   DoStatement       &     22.0,0.0,0.0 \\ 
 If       &  IfStatement       &    30.0,0.0,0.0 \\ 
  Switch   &   SwitchStatement   &     38.0,0.0,0.0 \\ 
  TernaryOp &   TernaryExpression &     39.0,0.0,0.0 \\ 
  While     &   WhileStatement    &     44.0,0.0,0.0 \\ 
\hline
\end{tabular}
\label{tab:CandJavaTokens}
\end{center}
\end{table}

\subsection{Generic Model generation}
\begin{figure}[htbp]
\centering
 	\centerline{\includegraphics[width=\linewidth]{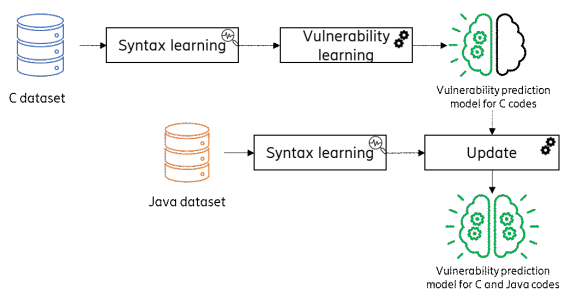}}
 	\caption{General approach used for vulnerability prediction knowledge transferring.}
	\label{fig:tl_general}
\end{figure}

In this study, we aim to overcome the problem of lack of sufficient training data samples to generalize an ML-model to predict a software vulnerability in a source code of a programming language (target). To this end, we train an ML-model for the same vulnerability type using a large number of data samples for another programming language and then transfer the knowledge for vulnerability prediction in the target language, by applying transfer learning method. 
The main approach we followed in this study (as shown in Figure ~\ref{fig:tl_general}) is based on two main steps: First, we trained and improved a model with a big dataset that includes vulnerability examples written in C. Then, we updated the model with the limited Java examples to predict the vulnerabilities existed in Java codes using the knowledge gained from C codes.  

To automatically detect the important features of vulnerabilities without the need for human intervention, we used CNN architecture. Another advantage of using CNN is to extract the relations and correlations between features by applying convolution between features.

\begin{figure}[htbp]
 	\centering
 	\includegraphics[width=\linewidth]{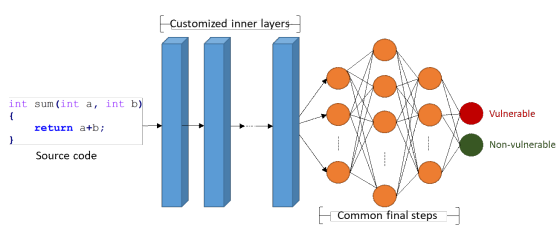}
 	\caption{Proposed model for vulnerability prediction knowledge transferring.}
  \vspace{-2em}
\label{fig:tl_model}
\end{figure}

For knowledge transferring, we developed an approach that is different from the common transfer learning approaches (where the input data in the two datasets are similar but the output labels are different). As in our case, we aim to get the same output labels (vulnerable and non-vulnerable) for Java codes leveraging the features and weights learned from C dataset. However, the input Java code instructions have structural and syntax differences and they need to be reflected on their equivalent C code instructions. Thus, as shown in Figure ~\ref{fig:tl_model}, we customized the first steps of our approach (including tokenization, AST generation, and encoding to numerical tuples) to process Java input codes and convert them into numerical arrays peer to the ones extracted from C codes. Then, we use the trained CNN for the feature extraction and classification steps. One of the advantages of this approach other than knowledge transferring, is that the trained feature extraction and classification layers can be used as they are, and they also can be improved and updated with new data samples (C or Java). 

Moreover, to decrease the impact of structural differences between the two programming languages (C and Java), we built the system to analyse source code on function level. 

\section{\uppercase{Dataset}}
\label{sec:dataset}
For this study, we used data samples from two datasets; Draper VDISC Dataset \cite{VDISC} and Software Assurance Reference Dataset (SARD) \cite{SARD}. While the former consists of the vulnerability samples for C source code, the latter includes vulnerabilities for Java source code. A brief summary related to each dataset is provided in the following subsections.
\subsection{Draper VDISC Dataset}
VDISC dataset involves function-level source code of C and C++ programming languages, labeled by static analysis for potential vulnerabilities. The data is provided in HDF5 files, and raw source code are stored as UTF-8 strings. Five binary 'vulnerability' labels are provided for each function, corresponding to the four most common CWEs as: CWE-120 (3.7\% of functions), CWE-119 (1.9\% of functions), CWE-469 (0.95\% of functions), CWE-476(0.21\% of functions), and CWE-other (2.7\%of functions).

\subsection{SARD Dataset}
SARD is a growing collection of test programs with documented weaknesses. Test cases vary from small synthetic programs to large applications. The source of the test cases in SARD dataset are: Wild code sampling (code from known bugs available in industry and open source software), artificial code constructing (codes produced by researchers to cover a wide range of weaknesses), academic code (Code collected from computer science and programming courses), and automatically generated codes. SARD dataset is provided in terms of test cases that provide the code files of each CWE type. In this paper we used Java source code test cases from Juliet Test Suite. 

\subsection{Dataset Preparation}
The evaluation of the model was carried out by analyzing NULL Pointer Dereference vulnerabilities (CWE-476), as it is a prevalent CWE that provides suitable data for both C and Java programming languages for the purposes of this study. For model training and validation, we obtained C data samples from the VDISC dataset, resulting in a prepared dataset of 120,609 non-vulnerable C functions and 1160 vulnerable C functions. However, the ratio of samples in the vulnerable class to the non-vulnerable class was observed to be significantly imbalanced, with more than 100 non-vulnerable samples for every 1 vulnerable sample, which impacts the model's ability to gain knowledge of vulnerable samples. To address this issue, we down-sampled the dataset to a ratio of 1:4. Additionally, we acquired Java samples from the SARD dataset, and after performing preprocessing steps, we generated a labeled dataset with 616 vulnerable Java functions and 328 non-vulnerable Java functions. Due to the limited number of Java samples in either class, we resorted to using transfer learning to apply the ML model trained on C codes to detect vulnerabilities in both C and Java codes. Finally, the labels were converted to a binary classification format, with 0 indicating non-vulnerable and 1 indicating vulnerable code sample.

\section{\uppercase{Results and Discussion}}
\label{sec:exp}
In this section, we present and discuss the performance result of the proposed method. First, we present the experimental results of the proposed model. Then, we discuss the validity, limitations, and implications of the new work. 

\subsection{Experimental Results}
We used a binary classifier based on a 1D CNN that consists of two convolutional layers, a maxpooling layer, dropout layers, a fully connected layer, and a softmax layer. The CNN architecture is chosen due to its ability to automatically detect the important features without the need for human intervention 
Also, by applying convolution between features, it extracts the relations and correlations between features. A dropout layer is applied to avoid over-fitting to the training data. 

Since we are training a binary classifier, it is better to train it with better balanced dataset. Thus, we used the down-sampled C dataset for the initial training of the CNN model. On the other hand, unbalanced test set represents the expected real world scenario where there are less functions that could induce vulnerabilities. Therefore, we use the full test set to measure the performance of the model. The sizes and the number of vulnerable and non-vulnerable samples used for training and test steps are shown in Table \ref{tab_train_test_samples}. As could be noted, Java related samples are few and not enough to train a model.   

\begin{table}[htbp]
\fontsize{6.32}{9}\selectfont
\caption{Number of vulnerable and non-vulnerable samples used in train and test sets.}
\begin{center}
\begin{tabular}{|l|c|c|c|c|}
\hline
\textbf{Data} & \textbf{Language}& \textbf{Down-sampled} & \textbf{Vulnerable}& \textbf{Non-vulnerable}\\ 
\hline
Train   & C  & Yes  & 1160    & 4640\\
Train   & Java  & No  & 305   & 167\\
Test    & C    & No   & 140  & 15,062\\
Test    & Java  & No & 311 & 161\\
\hline
\end{tabular}
\label{tab_train_test_samples}
\end{center}
\end{table}
To assess the performance of proposed model in vulnerabilities predicting, we analyze accuracy, precision, recall, F1-score, and Area Under the Curve (AUC). The confusion table on both C and Java test sets is presented in Table \ref{tab:cms}. And the evaluating metrics both C and Java source code are provided in Table \ref{results_c_and_Java}. As could be observed, the developed model is able to detect vulnerabilities in both C and Java source code with accuracy of 99\% and 91\% respectively. Hence, we transferred the knowledge the model gained from C vulnerable samples to be used to also detect vulnerabilities in Java source code. Moreover, the additional Java samples used to update the model helped to improve the model, as Precision, Recall, F1-score, and AUC percentages have been increased.

\begin{table}[b!]
\fontsize{8}{9}\selectfont
\caption{Confusion tables for C and Java examples (No = non-vulnerable, and Yes = vulnerable).}\centering
\begin{center}
\begin{tabular}{|clcc|} 
\hline
\multicolumn{1}{|}{}  &     & \multicolumn{2}{c|}{~Predictions}  \\ 
\hline
\multicolumn{2}{|c|}{\textbf{C test set}}          & No & Yes\\ 
\hline
\multirow{2}{*}{Actuals} & \multicolumn{1}{|l|}{No} & 14977          & 85                \\
                         & \multicolumn{1}{|l|}{Yes}     & 64             & 76                \\ 
\hline
\multicolumn{2}{|c|}{\textbf{Java test set}}       &No&Yes\\ 
\hline
\multirow{2}{*}{Actuals} & \multicolumn{1}{|l|}{No} & 128            & 33                \\
                         & \multicolumn{1}{|l|}{Yes}     & 10             & 301               \\
\hline
\end{tabular}
\label{tab:cms}
\end{center}
\end{table}

\begin{table}[t!]
\fontsize{8}{9}\selectfont
\caption{Vulnerability prediction results for both C and Java samples.}\centering
\begin{center}
\begin{tabular}{|l|c|c|c|} 
\hline
\textbf{Performance} & \textbf{C} & \textbf{Java} & \textbf{Average}  \\ 
\hline
Accuracy                            & 0.99       & 0.91          & 0.95              \\
Precision                           & 0.47       & 0.90          & 0.68              \\
Recall                              & 0.54       & 0.97          & 0.75              \\
F1-Score                            & 0.51       & 0.93          & 0.72              \\
AUC                                 & 0.77       & 0.88          & 0.82              \\
\hline
\end{tabular}
\label{results_c_and_Java}
\end{center}
\end{table}

\subsection{Model Explanation}
\begin{figure}[htbp]
  \centering
 	\includegraphics[width=\linewidth,height= 1.5in]{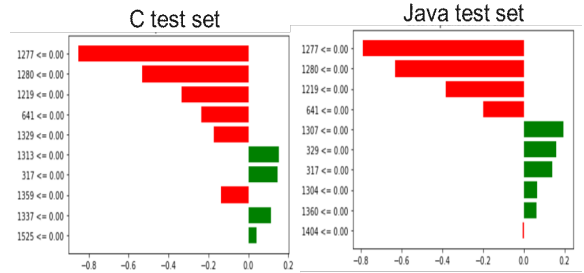}
 	\caption{Local explanation of vulnerable examples detection in C and Java test sets.}
	\label{fig:lime}
\end{figure}

We utilize Lime (Local Interpretable Model-agnostic Explanations) method to explain and interpret the predictions of our proposed model. We aim to examine the effect of our code representation method on the classification model for each programming language. For this aim, we apply Lime on both C and Java test sets separately, and analyze the features used to differentiate vulnerable source code for each language. Analysing results of Lime are presented in Figure ~\ref{fig:lime}. 
We can notice that the most effective features (encoded values) used for differentiating vulnerable from non-vulnerable source code are the same for both programming languages. For example, the value 1277, 1280, 1219, 641, and 317 have similar effect on Java and C source code. This proves that our method have successfully converted the source code from both programming languages (C and Java) into unique, equivalent, simple, ML friendly numeric vector. In addition, it shows that proposed model is generic and can be used to detect trained vulnerabilities in other programming languages once same code representation method is applied. 

\subsection{Threats to Validity}
\textbf{Data Insufficiency:} The limited availability of data for both C and Java source code may impact the representativeness of the results. To address this, data from two different datasets were used. However, this could lead to potential biases in the results if the data sets are not fully representative of real-world applications. \textbf{Preprocessing:} The preprocessing steps involved matching the syntax between C and Java tokens, which may have introduced errors or biases due to the structural differences between the two languages. \textbf{Model Training:} To address the lack of labeled source code from real projects, synthetic codes from the SARD and VDISC datasets were used for training the model. This could affect the validity of the results, as the model may not generalize well to real-world applications.
\section{\uppercase{Conclusion and future work}}
In this study, we focus on the problem of having vulnerability samples for some programming languages but not others. To overcome this problem, we design a method that extract vulnerability prediction knowledge from available data samples and then use it to predict vulnerabilities in another programming language. We also, add flexibility to update the model once new samples are provided. Specifically, in this study, we built a model that is able to detect vulnerabilities in both Java and C source code. We trained a CNN-based model with C source code from VDISC dataset. Then, we modified the model to detect the learned vulnerabilities in Java source code. We extracted Java sample codes from SARD dataset. By the end of our experiments, we were able to show that despite the many differences between programming languages, we were able to train one model to detect vulnerabilities in more than one programming language. This study could be further extended to detect vulnerabilities in other commonly used programming languages such as Python and Javascript. The study could be also improved by training the model on other common vulnerability types from different programming languages. 
\section*{\uppercase{Acknowledgements}}
This work was funded by The Scientific and Technological Research Council of Turkey, under 1515 Frontier R\&D Laboratories Support Program with project no: 5169902.
\bibliographystyle{apalike}
{\small
}

\end{document}